\documentclass{aa}  
\usepackage{graphicx}
\usepackage{natbib}
\usepackage{longtable}
\usepackage{rotating}
\usepackage[varg]{txfonts}
\usepackage{pdflscape}
\usepackage{multirow}

\def\degr{\hbox{$^\circ$}}
\def\arcmin{\hbox{$^\prime$}}
\def\arcsec{\hbox{$^{\prime\prime}$}}

\newcommand{\nclouds}{11303 }

\newcommand{\microns}{$\mu$m }

\usepackage{txfonts}
\begin{document}
   \title{The initial conditions of stellar protocluster formation}

   \subtitle{I. A catalogue of Spitzer dark clouds}

   \author{N. Peretto
          \inst{1}
          \and
          G.~A. Fuller\inst{1}
          }

   \offprints{}

   \institute{Jodrell Bank Centre for Astrophysics, Alan Turing
     Building, School of Physics and Astronomy, The University of Manchester, Oxford Road,  Manchester M13 9PL, UK\\
              \email{Nicolas.Peretto@manchester.ac.uk}
                      }

   \date{Received; accepted}

 
   \abstract
  {The majority of stars form in clusters. Therefore a comprehensive
    view of star formation requires understanding the initial
    conditions for cluster formation. }
  {The goal of our study is to shed light on the physical properties of
    infrared dark clouds (IRDCs) and the role they play in the formation of
    stellar clusters. This article, the first of a series dedicated to the
    study of IRDCs, describes techniques developed to establish a complete
    catalogue of Spitzer IRDCs in the Galaxy.  
     }
     { We have analysed Spitzer GLIMPSE and MIPSGAL data to identify a
       complete sample of IRDCs in the region of Galactic longitude and
       latitude $10^\circ < |l|<65^\circ$ and $|b|<1^\circ$. From the 8$\mu$m
       observations we have constructed opacity maps and used a newly
       developed extraction algorithm to identify structures above a column
       density of N$_{\rm{H_2}} \ga 1\times10^{22}$~cm$^{-2}$. The
       24$\mu$m data are then used to characterize the star formation
       activity of each extracted cloud. }
     { A total of 11303 clouds have been extracted. A comparison with the
       existing MSX based catalogue of IRDCs shows that 
         80$\%$ of these Spitzer dark clouds were previously unknown.
       The algorithm also
       extracts $\sim$ 20000 to 50000 fragments within these clouds, depending
       on detection threshold used. 
       A first look at the MIPSGAL data indicates that between 20\% and 68\% of these IRDCs show
       24\microns
       point-like association.
       This new database provides an
       important resource for future studies aiming to understand the initial
       conditions of star formation in the Galaxy.}
{}


   \maketitle
%

\section{Introduction}

\begin{figure*}[!t!]
\includegraphics[width=6cm,angle=270]{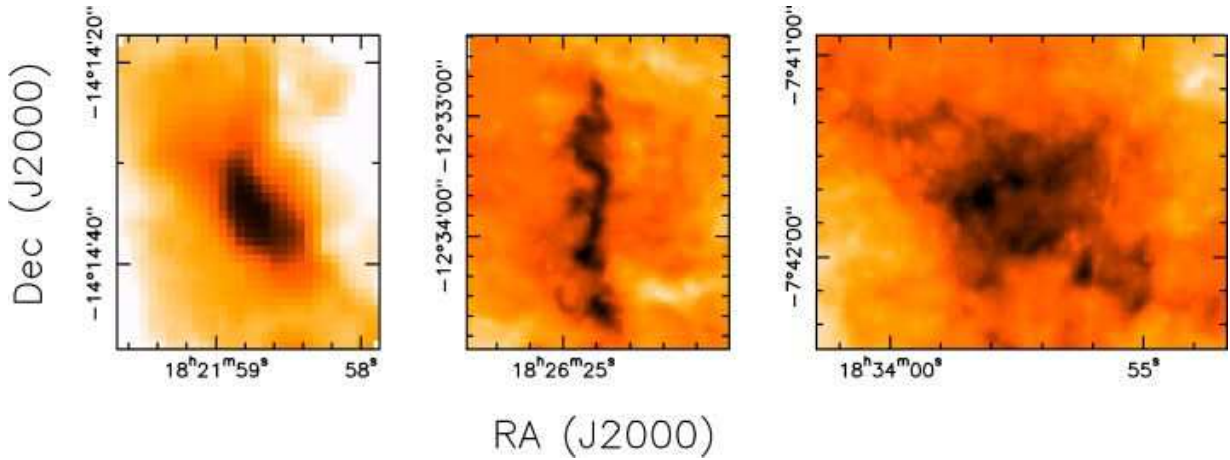}
\caption{ These images show the GLIMPSE Spitzer 8$\mu$m emission of 3
  random IRDCs  from our sample. These illustrate the diversity in shape and size of IRDCs.
  \label{irdc_samp}}
\end{figure*}

The majority of stars form in groups from few tens to few hundreds of objects
\citep[e.g.][]{ladalada2003}. So, understanding cluster formation is key to
understanding the formation of stars.  Clusters form from the gas located in
the densest parts of molecular clouds, within structures called clumps
\citep[]{blitz1993}. These clumps fragment into an assembly of protostellar
cores which collapse to produce stars, forming `protoclusters'. By definition,
protoclusters are active star forming regions, with jets, flows
and heating sources \citep[e.g.][]{bally2006} which rapidly start to shape
their surroundings.
From the study of these protoclusters, it is therefore difficult to back track
to the initial conditions of their formation.  On the other hand, clumps which
are on the verge of forming protostars, but which have not formed any yet, are
structures unpolluted by star formation activity and must still reflect the
initial conditions of the formation of
protoclusters. Looking for, and studying such
`pre-protoclusters' is crucial for our understanding of star formation
processes.

Only a tiny percentage of the material in any molecular cloud forms stars.
These star-forming regions are traced by various signposts of star formation
activity such as the presence of strong infrared sources, outflows, jets, 
  methanol and water masers and compact HII regions.  The problem with
identifying pre-protoclusters is that by definition these signposts are not
yet present.  Other means are thus necessary to find such objects. The two
infrared satellites ISO and MSX have been important tools for this
purpose. Indeed, the large infrared surveys these satellites carried out
identified infrared dark structures, seen in absorption from 7 to 25 $\mu$m
against the background emission \citep{perault1996, hennebelle2001, egan1998,
  simon2006a} .  Millimeter molecular lines \citep[e.g.][]{carey1998,
  teyssier2002, pillai2006} and dust continuum observations
\citep[e.g.][]{teyssier2002, rathborne2006} have clearly demonstrated that
these infrared dark clouds are dense, cold structures, possibly being the
progenitors of protoclusters \citep{simon2006b}.  \citet{rathborne2006} even
suggested that the dust continuum ``cores" observed in these IRDCs are the
direct progenitors of massive stars.  However, the wide range of mass and size
of these IRDCs clearly suggests that they cannot all be evolving along the
same evolutionary path and they must lead to the formation of a large range of
different stellar contents.

So far, the study of the earliest stages of the formation of
protoclusters have mostly focussed on the closest objects such as
$\rho$-Oph\citep[e.g.][]{motte1998, andre2007}, Perseus
\citep{hatchell2005, enoch2006}, NGC2264
\citep[e.g.][]{peretto2006, teixeira2006}.  The results of these
studies set important constraints on models of star formation, but may
not be representative of the formation of stars throughout the Galaxy.
The only way to define such a representative view is through
studies of large unbiased samples of the precursors of stellar
clusters.

In this paper we identify and characterise the IRDCs detected using the Spitzer
GLIMPSE and MIPSGAL archive data. The high angular resolution of the Spitzer
data provides a detailed probe of the structure of these sources while the
high sensitivity of IRAC and MIPS allows us to detect previously unseen deeply
embedded protostars/protoclusters.  Section 2 of this paper presents the
Spitzer archive data used for this study. Section 3 will discuss the
construction of 8$\mu$m opacity maps for IRDCs, while Section 4 will focus on
the conversion from 8\microns opacity to H$_2$ column density. The extraction
of structures within these maps will be discussed in Section 5. A comparison
with the MSX catalogue of IRDCs is in Section 6 while Section 7 summarizes our
initial study.  The nature of these dark clouds and their star formation
actively are discussed in more detail in subsequent papers (Peretto \& Fuller,
in preparation).


\section{A large survey of infrared dark clouds: Spitzer archive data}

IRDCs are seen in silhouette against the infrared background emission (see
Fig.~\ref{irdc_samp}) and as a sample are likely to contain protoclusters and
pre-protoclusters. Even when large scale (sub)millimetre surveys of the
Galactic plane become available and these objects can be detected through
their dust emission, IRDCs and studies of the absorption towards these sources
will remain important. Not only can the IRDCs be studied at high angular
resolution at infrared wavelengths, but unlike the (sub)millimetre emission,
their column density can be measured from the absorption independent of the
dust temperature.

The first large survey of IRDCs was undertaken by \citet{simon2006a} using the
mid-infrared data of the MSX satellite. In total, Simon et al.  detected more
than 10000 IRDCs, with sizes larger than (36\arcsec)$^2$ and flux density more
than 2~MJy/sr ($> 2$ times the rms noise of the MSX images) below the
mid-infrared radiation field. Within these IRDCs they extracted more than
12000 IRDC ``cores".  \citet{simon2006b} performed a follow up of a sub-sample
of few hundreds sources for which they were able to determine distances. They
found that these IRDCs are very similar to CO molecular clumps  \citep[e.g.][]{blitz1993}.

In the GLIMPSE and MIPSGAL surveys the Spitzer satellite has resurveyed a
large fraction of the Galactic plane at infrared wavelengths ($10\degr <|l|<
65\degr, 0<|b|< 1\degr$).  These data have both better angular resolution
(2\arcsec~ vs 20\arcsec\ at 8\microns) and sensitivity (0.3~MJy/sr vs
1.2~MJy/sr at 8\microns) than the MSX data, as well as wider wavelength
coverage.The IRAC (3.6, 4.5, 5.8, 8 $\mu$m) GLIMPSE and MIPS (24, 70, 160
$\mu$m) MIPSGAL observations provide a unique opportunity to shed light on
the role of IRDCs during the earliest stages of star formation.
Despite a smaller coverage of the Galactic plane by Spitzer, 
an initial comparison of the MSX IRDC catalogues with the Spitzer
observations indicated that the Spitzer data contained IRDCs undetected by MSX 
 in the same region of the
Galaxy. Therefore an unbiased search of the Spitzer GLIMPSE data has
been undertaken to identify IRDCs.


Many IRDCs can been seen in silhouette up to at least 24 $\mu$m, providing a
wide wavelength range over which they can be studied in absorption. However
several factors affect the choice of the optimal wavelength at which to
identify and study the overall cloud properties. These include the strength
and uniformity of the background emission and the number of foreground and
background stars and in principle, the wavelength dependence of the dust
extinction law, although recent work suggests that from 4.5 to 8$\mu$m,
  the three last bands observed by Spitzer/IRAC, the extinction is a
relatively flat function of wavelength
\citep{lutz1996,indebetouw2005,roman-zuniga2007}.  The angular resolution of
the observations is highest at the shortest wavelengths, but in these bands a
very high density of stars is detected and high degree of structure in the
relatively weak background emission makes analysis of the images at these
wavelengths complex.  Overall, inspection of the Spitzer data shows that the
strength and relative smoothness of the background emission together with the
relatively low density of stars make the IRAC 8 $\mu$m band the most suitable
for this initial study of a large sample of objects.

The GLIMPSE and MIPSGAL data have been reduced and calibrated automatically to
produce the so called post-Basic Calibrated Data (post BCD). The typical flux
uncertainty for point-like sources is $\sim2\%$ at 8$\mu$m \citep{reach2005}
while the position uncertainty is less than 0.3\arcsec (IRAC manual V8.0:
http://ssc.spitzer.caltech.edu/documents/SOM/). However, because we are not
looking at point-like sources but extended objects, a calibration factor has
to be applied on the PBCD 8$\mu$m images \citep{reach2005}. This calibration
factor, CF, is a function of the aperture radius, R$_a$, for the source under
investigation (http://ssc.spitzer.caltech.edu/irac/calib/extcal/). The
relation between CF and R$_a$ in arcseconds, at 8$\mu$m is $CF = 1.37\times
\exp(-R_a^{0.33}) + 0.74$. Because the typical size of the structure we
analyse is about one arcminute, in the analysis which follows we applied a
calibration factor of 0.8 to the PBCD 8$\mu$m images. A different
calibration factor would not change the opacities of the IRDCs we calculated, but would imply different  related intensities (Table \ref{table:1}).

\begin{figure}[!t!]
\includegraphics[width=6cm,angle=270]{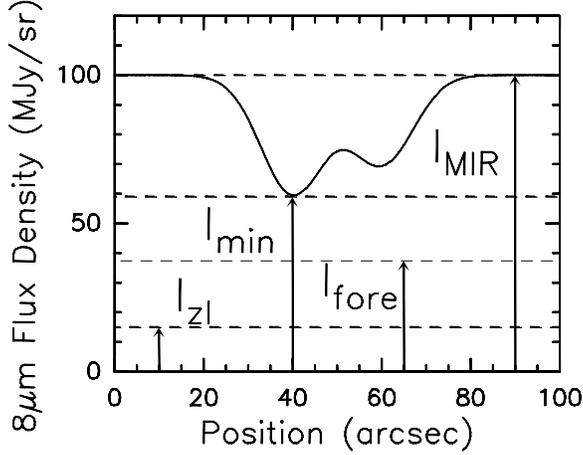}
\caption{Schematic view a typical IRDC flux density profile. The variable meanings used in the rest of the text are illustrated on this figure. In this figure, $I_{\rm{fore}}$ has been set to a particular value, e.g., 38~MJy/sr, but in practice, it can be anywhere between $I_{\rm{zl}}$ and $I_{\rm{min}}$.
  \label{scheme_irdc}}
\end{figure}

\section{Opacity distribution of IRDCs}

\subsection{Principle}

Infrared dark clouds are structures seen in absorption against the background
emission. The strength of the absorption is directly related to the opacity
along the line of sight. Following the notation of \citet{bacmann2000}, the relation between the opacity $\tau_{\lambda}$ and
the intensity at wavelength at $\lambda$, emerging from the cloud
$I_{\lambda}$, is given by
\begin{equation}
  I_{8\mu\rm{m}} = I_{\rm{bg}-8\mu\rm{m}}\times \exp(-\tau_{8\mu\rm{m}}) + I_{\rm{fore}-8\mu\rm{m}}
\label{eq1}
\end{equation}
where $I_{\rm{bg}-8\mu\rm{m}}$ is the intensity of the background
emission at 8$\mu$m, and $I_{\rm{fore}-8\mu\rm{m}}$ is the foreground
emission. In the following for simplicity we drop the 8$\mu$m
  label on the variable names, except on the opacity. If we know the
foreground and background intensities we can invert Eq.~(\ref{eq1})
and infer the spatial distribution of the opacity within an infrared
dark cloud,
\begin{equation}
  \tau_{8\mu\rm{m}} = - \ln\left(\frac{I - I_{\rm{fore}}}{I_{\rm{bg}}}\right)
\label{eq2}
\end{equation}

$I_{\rm{fore}}$ and $I_{\rm{bg}}$ are related to each other by $I_{\rm{MIR}} =
I_{\rm{bg}} + I_{\rm{fore}}$ where $I_{\rm{MIR}}$ is the observed mid-infrared
radiation field and can be estimated directly from the $8\mu$m images (see
Fig.~\ref{scheme_irdc}).  A lower limit on $I_{\rm{fore}}$ is given by the
intensity of the zodiacal light, $I_{\rm{zl}}$, in the direction of the cloud,
while an upper limit is given by the minimum intensity within the cloud,
$I_{\rm{min}}$.  However, with the extinction data only, it is impossible to
find the exact value of $I_{\rm{fore}}$ for a given cloud.

\begin{figure}[!t!]
\hspace{0.5cm}
\includegraphics[width=6.0cm,angle=270]{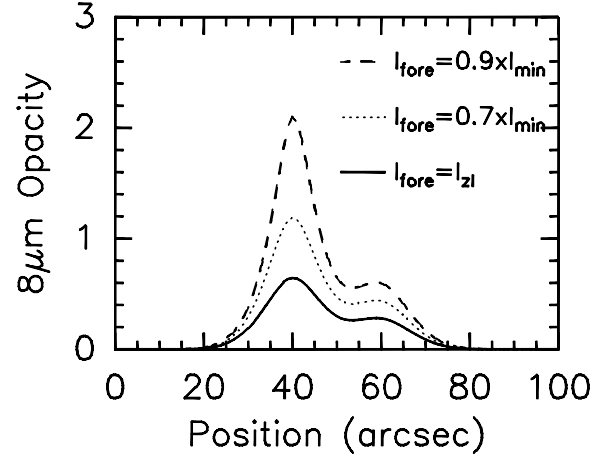}
\caption{Calculated opacity profiles of the IRDC plotted in
  Fig.\ref{scheme_irdc} corresponding to 3 different assumptions on the
  foreground intensity. The solid line shows $I_{\rm{fore}}=I_{\rm{zl}}$ (i.e
  $I_{\rm{fore}}=0.25\times I_{\rm{min}}$), the dotted line $
  I_{\rm{fore}}=0.7\times I_{\rm{min}}$ and the dashed line $
  I_{\rm{fore}}=0.9\times I_{\rm{min}}$ . }\label{plot_err}
\end{figure}

The determination of $I_{\rm{fore}}$ is crucial to infer the spatial opacity
distribution of a given IRDC. To illustrate this point, we computed the
  opacity of the cloud profile shown in Fig.~\ref{scheme_irdc} for three
  different values of $I_{\rm{fore}}$ (Fig.~\ref{plot_err}). On this figure we
  see that, with increasing $I_{\rm{fore}}$, the opacity increases
  significantly everywhere in the cloud, and even more sharply at the
  peak. These opacity variations are even more drastic for shallower clouds. 
It is therefore important to constrain
$I_{\rm{fore}}$ when calculating the opacity distribution of an IRDC.

Of course it is also possible that at least some the IRDCs are saturated and
their intensity profiles become flattened. In such cases, it becomes
impossible to recover the central structure of the clouds through the
extinction maps.  Moreover, such flattening could lead to an incorrect
interpretation of the final opacity profiles of IRDCs. 

\begin{sidewaystable*}[!h!]
\caption{SDC properties for the first 30 out of \nclouds in the catalogue. The
  full catalogue is available online. The columns give a running number (1),
  the name of the source based in its Galactic coordinates (2), the right
  ascension and declination (in J2000) of the cloud peak (3,4), the minimum
  8$\mu$m emission towards the cloud ($I_{\rm{min}}$) (5), the background
  8$\mu$m emission ($I_{\rm{MIR}}$) (6), the maximum $I_{\rm{MIR}}$ variation
  within the IRDC ($\delta I_{\rm{MIR}}$, Sec.~3.3) (7), the size of the
  cloud along its major and minor axes in arcseconds (8,9), the position angle
  of the major axis of the cloud in degrees East of North (10), the equivalent
  radius ($R_{\rm{eq}}$; Sec.~\ref{sec:irdcs}) of the cloud (11), the peak and
  average optical depth of the cloud at 8$\mu$m (12,13), the optical depth at
  8$\mu$m at which the absorption would be saturated (14), the number of
  fragments in the cloud identified with $\tau_{\rm{step}}=0.35$ (15;
  Sec.~5.2), whether there are 24\microns stars in the field (16) and in the
  cloud (17; Sec.~\ref{sec:stars}); and the density of stars around the cloud
  (18).  }
\label{table:1}
\centering
\begin{tabular}{cccccccccccccccccc}
\hline
\hline
 Number & Name & \multicolumn{2}{c}{Coordinates}  & $I_{\rm{min}}$ &
$I_{\rm{MIR}}$& $\delta I_{\rm{MIR}}$  & $\Delta X$ &  $\Delta Y$ &
$\alpha$ &  R$_{\rm eq}$& $\tau_{\rm{peak}}$ & $\tau_{\rm{av}}$ & $\tau_{\rm{sat}}$ & frag & \multicolumn{2}{c}{star} & $\sigma_{\rm{star}}$\\
& & RA(J2000) & Dec(J2000)  & (MJy/sr) &(MJy/sr)   &  & (\arcsec) &  (\arcsec) & (\degr) &  (\arcsec)&  &  &  &  &Field & IRDC &(arcmin$^{-2}$) \\
\hline
 1 &        SDC9.22+0.169 & 18:05:30.40 & -20:53:16.0 &   37.9 &   55.7 & 0.07 &   42.9 &   29.8 & 76 &   32.3 &  1.16 &  0.50 &  4.45 &   2 & y & y &  0.92 \\
 2 &        SDC9.25+0.144 & 18:05:39.69 & -20:52:25.6 &   40.8 &   56.9 & 0.04 &   26.9 &   20.5 & -66 &   23.9 &  0.95 &  0.48 &  4.47 &   1 & y & n &  0.95 \\
 3 &       SDC9.256+0.133 & 18:05:42.92 & -20:52:27.8 &   41.7 &   56.7 & 0.01 &   13.5 &   10.8 & -42 &   13.1 &  0.85 &  0.48 &  4.47 &   1 & n & n &  1.33 \\
 4 &       SDC9.301+0.126 & 18:05:50.01 & -20:50:16.8 &   43.0 &   56.7 & 0.00 &   17.9 &    7.2 & -89 &   12.6 &  0.75 &  0.45 &  4.47 &   1 & n & n &  1.34 \\
 5 &       SDC9.328+0.031 & 18:06:14.76 & -20:51:39.7 &   46.2 &   60.8 & 0.02 &   11.4 &   10.0 & 87 &   11.6 &  0.74 &  0.47 &  4.53 &   1 & y & n &  1.14 \\
 6 &        SDC9.403+0.26 & 18:05:33.02 & -20:41:00.7 &   33.7 &   44.5 & 0.02 &   11.2 &    7.3 & -87 &   10.2 &  0.75 &  0.44 &  4.22 &   1 & y & n &  0.93 \\
 7 &       SDC9.432+0.163 & 18:05:58.31 & -20:42:21.8 &   38.7 &   51.4 & 0.03 &   13.3 &    8.4 & -50 &   11.8 &  0.76 &  0.44 &  4.37 &   1 & n & n &  1.02 \\
 8 &       SDC9.461+0.138 & 18:06:07.50 & -20:41:32.8 &   40.0 &   53.0 & 0.03 &   17.8 &    6.9 & 22 &   12.0 &  0.77 &  0.42 &  4.40 &   1 & n & n &  1.08 \\
 9 &       SDC9.624+0.187 & 18:06:16.80 & -20:31:35.6 &   37.7 &   53.5 & 1.58 &  321.6 &  203.8 & -75 &  188.7 &  4.30 &  0.61 &  4.41 &   6 & y & y &  0.78 \\
10 &       SDC9.629-0.061 & 18:07:13.25 & -20:38:38.7 &   38.0 &   51.7 & 0.03 &   23.0 &   16.1 & 30 &   18.0 &  0.85 &  0.45 &  4.37 &   1 & y & n &  0.94 \\
11 &       SDC9.635+0.296 & 18:05:53.87 & -20:27:49.8 &   24.1 &   39.7 & 0.08 &   29.6 &   15.1 & -60 &   21.5 &  1.96 &  0.58 &  4.11 &   1 & y & n &  0.76 \\
12 &       SDC9.689+0.000 & 18:07:06.92 & -20:33:41.5 &   39.6 &   52.9 & 0.01 &   24.6 &   11.8 & 57 &   15.9 &  0.79 &  0.46 &  4.40 &   1 & n & n &  1.01 \\
13 &        SDC9.692-0.55 & 18:09:10.79 & -20:49:34.6 &   27.2 &   35.3 & 0.02 &   18.6 &    6.8 & -69 &   12.7 &  0.69 &  0.41 &  3.99 &   1 & n & n &  0.94 \\
14 &       SDC9.737-0.239 & 18:08:06.57 & -20:38:08.1 &   37.9 &   52.1 & 0.05 &   30.4 &   12.4 & 4 &   20.3 &  0.91 &  0.48 &  4.38 &   1 & n & n &  0.97 \\
15 &       SDC9.762-0.567 & 18:09:23.29 & -20:46:20.9 &   26.6 &   37.3 & 0.01 &   12.1 &    7.5 & -48 &   10.7 &  0.98 &  0.51 &  4.05 &   1 & y & y &  0.79 \\
16 &       SDC9.787-0.156 & 18:07:54.17 & -20:33:06.8 &   40.1 &   61.2 & 0.06 &   58.7 &   26.3 & -26 &   37.1 &  1.38 &  0.57 &  4.54 &   2 & y & n &  1.13 \\
17 &       SDC9.796-0.028 & 18:07:26.89 & -20:28:53.3 &   43.5 &   59.8 & 0.04 &   36.6 &   17.7 & -1 &   25.7 &  0.92 &  0.50 &  4.52 &   1 & n & n &  1.25 \\
18 &       SDC9.798-0.707 & 18:09:59.42 & -20:48:32.9 &   29.1 &   40.5 & 0.36 &   63.7 &   45.8 & 2 &   47.5 &  0.90 &  0.47 &  4.13 &   1 & y & n &  0.72 \\
19 &       SDC9.819-0.141 & 18:07:54.93 & -20:31:00.7 &   44.8 &   62.2 & 0.02 &   39.0 &   12.5 & 37 &   20.1 &  0.95 &  0.44 &  4.56 &   1 & y & n &  1.25 \\
20 &        SDC9.825-0.03 & 18:07:30.72 & -20:27:26.0 &   41.0 &   58.3 & 0.08 &   78.9 &   19.9 & 78 &   32.7 &  1.02 &  0.47 &  4.49 &   2 & y & n &  1.16 \\
21 &       SDC9.844+0.752 & 18:04:38.31 & -20:03:31.4 &   23.1 &   30.8 & 0.05 &   24.1 &   13.6 & -52 &   16.8 &  0.76 &  0.43 &  3.86 &   1 & y & n &  0.56 \\
22 &       SDC9.845-0.138 & 18:07:57.47 & -20:29:34.3 &   37.0 &   63.5 & 0.05 &   67.9 &   35.1 & 0 &   37.5 &  2.28 &  0.51 &  4.58 &   2 & y & n &  1.16 \\
23 &       SDC9.852-0.034 & 18:07:35.07 & -20:26:07.8 &   30.5 &   59.3 & 0.17 &  119.4 &   55.6 & -73 &   76.5 &  4.95 &  0.87 &  4.51 &  12 & y & y &  1.19 \\
24 &       SDC9.859-0.746 & 18:10:15.75 & -20:46:27.5 &   33.1 &   61.3 & 1.94 &  294.5 &  150.3 & -7 &  162.8 &  3.08 &  0.79 &  4.54 &   4 & y & y &  0.63 \\
25 &       SDC9.864-0.102 & 18:07:51.78 & -20:27:30.2 &   49.1 &   64.5 & 0.02 &   35.2 &   17.3 & 84 &   23.4 &  0.74 &  0.44 &  4.59 &   1 & y & n &  1.18 \\
26 &       SDC9.872-0.767 & 18:10:22.02 & -20:46:23.9 &   42.5 &   69.3 & 1.51 &  180.5 &  113.6 & -66 &  126.3 &  2.29 &  0.79 &  4.67 &   2 & y & y &  0.55 \\
27 &        SDC9.878-0.11 & 18:07:55.37 & -20:26:58.8 &   35.5 &   65.7 & 0.07 &   59.2 &   40.9 & 26 &   49.3 &  4.98 &  0.83 &  4.61 &   4 & y & y &  1.18 \\
28 &       SDC9.889-0.747 & 18:10:19.58 & -20:44:56.8 &   61.7 &   99.2 & 0.20 &   26.5 &   23.2 & -16 &   24.8 &  2.01 &  0.92 &  5.02 &   1 & y & n &  0.59 \\
29 &       SDC9.895-0.749 & 18:10:20.74 & -20:44:41.4 &   73.5 &  105.0 & 0.04 &   12.6 &    6.4 & 18 &    9.9 &  1.07 &  0.54 &  5.08 &   1 & n & n &  0.57 \\
30 &       SDC9.904-0.699 & 18:10:10.70 & -20:42:43.9 &   58.7 &   78.9 & 0.38 &   39.5 &   19.2 & 43 &   26.9 &  0.73 &  0.47 &  4.80 &   1 & y & n &  0.49 \\
\hline
\end{tabular}
\end{sidewaystable*}

\subsection{Constraining $I_{\rm{fore}}$
\label{sec:ifore}}

\begin{figure}[!t!]
\includegraphics[width=9.cm,angle=270]{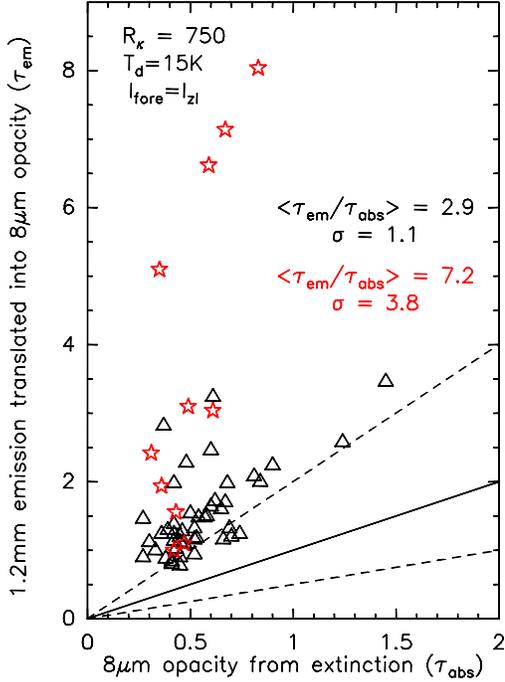}
\caption{Plot of the 8$\mu$m opacity estimated from the 8$\mu$m
  Spitzer maps ($\tau_{\rm{abs}}$) and from the 1.2mm dust continuum
  emission ($\tau_{\rm{em}}$). The starless sources are marked with open triangles while
  those associated with 24$\mu$m point-like emission are marked with red open star symbols. $\tau_{\rm{abs}}$ has been calculated assuming $I_{\rm{fore}}=I_{\rm{zl}}$.
  The solid line marks the relationship $\tau_{\rm{abs}} = \tau_{\rm{em}}$,
  while the two dashed lines indicates $\tau_{\rm{abs}} = 0.5\times\tau_{\rm{em}}$ and
  $\tau_{\rm{abs}} = 2\times\tau_{\rm{em}}$
  \label{rath}}
\end{figure}

Comparison of the infrared extinction and millimeter emission can be used to
constrain the infrared foreground emission towards an IRDC by requiring that
both techniques give the same column density towards the source.
For this purpose we have used the 38 IRDC 1.2mm dust continuum images
\cite{rathborne2006} obtained with the IRAM 30m telescope at 11\arcsec angular
resolution.  The 1.2mm emission can be translated into an 8$\mu$m opacity,
$\tau_{\rm{em}}$, using the equation

\begin{equation}
\tau_{\rm{em}} = \frac{S_{\rm{peak}} \times R_{\kappa}}{B_{1.2}(T_d)\times \Omega_{\rm{30m}}}
\end{equation}
where S$_{\rm{peak}}$ is the 1.2mm dust continuum emission peak of the source,
$R_{\kappa}$ is the specific dust opacity ratio between 8$\mu$m and 1.2mm,
$B_{1.2}(T_d)$ is the Planck function at 1.2mm for the dust temperature $T_d$,
and $\Omega_{30m}$ is the solid angle at 1.2mm of the IRAM 30m telescope
beam. The value for $R_{\kappa}$ is not well constrained: different models of
dusts provide different values of $R_{\kappa}$. Given the chemical composition
of the emitting/absorbing dust the value of $R_{\kappa}$ can be as large as
2000 for interstellar dust in diffuse clouds \citep[e.g.][]{draine2003},
decreasing to 750 for dense clouds
\citep[e.g.][]{ossenkopf1994,johnstone2003}. Given the dense and cold nature
of IRDCs, we adopted the value $R_{\kappa}=750$, and a dust temperature of
15~K, which gives
\begin{equation}
\tau_{\rm{em}} = 0.02 \times S_{\rm{peak}}
\end{equation}
with $S_{\rm{peak}}$ in mJy/beam.  After smoothing the Spitzer 8\microns\
images of the 38 IRDCs observed by \citet{rathborne2006} to the same resolution as the the IRAM 30m 1.2mm images, we have constructed their 8\microns\ opacity maps assuming $I_{\rm{fore}}=I_{\rm{zl}}$ (i.e. the lower limit on the foreground emission). A direct comparison between these opacity maps and the ones calculated from the 1.2mm dust continuum images becomes then possible. 
However the observations of the 8\microns absorption and 1.2mm emission are
not equally sensitive to all of the dust along the line of sight. 
Regions of low column density are more easily detected in
absorption than in emission.
For this reason, we selected only clear
corresponding peaks in both type of images, ending up with 57 ``cores"
(emission peaks and absorption minima) which have been used for the
comparison.  Amongst these cores 11 show 24$\mu$m point-like emission.
Figure \ref{rath} shows the resulting comparison for these 57 cores, the
``starless" ones (those without associated 24\microns\ emission) are marked
with open triangles while the ``protostellar" ones are marked with red stars.
Also shown are the three lines: $\tau_{\rm{abs}} = \tau_{\rm{em}}$ (solid
line), $\tau_{\rm{abs}} = 2\times \tau_{\rm{em}}$, and $\tau_{\rm{abs}} =
0.5\times \tau_{\rm{em}}$ (dashed lines).  In the figure there is a clear
separation between the starless sources and those objects associated with a
24$\mu$m point-like source.  For the sources associated with 24$\mu$m
point-like emission, the values of $\tau_{\rm{em}}$ are on average higher than
for the starless sources. The $\tau_{\rm{em}}/\tau_{\rm{abs}}$ ratio is on
average $\sim2.9$ for the starless sources with a dispersion of 1.1, while it
is $\sim7.2$ for the sources with stars with a dispersion of 3.8.  This
reflects that the latter group of sources have stronger 1.2mm emission (a
factor of $\sim2.5$ ), which translates to higher opacities for the same
assumed dust temperature. This clearly shows these sources are in fact either
warmer with average dust temperature greater than 15~K, or else have different
dust properties.
On the other hand for the starless objects, the average ratio
$<\tau_{\rm{em}}/\tau_{\rm{abs}}> =2.9$ is closer, but still rather far from,
unity.  This suggests that the value of $I_{\rm{fore}}$ is underestimated and
the assumption $I_{\rm{fore}}=I_{\rm{zl}}$ is incorrect. 

\begin{figure}[!t!]
  \includegraphics[width=6.cm,angle=270]{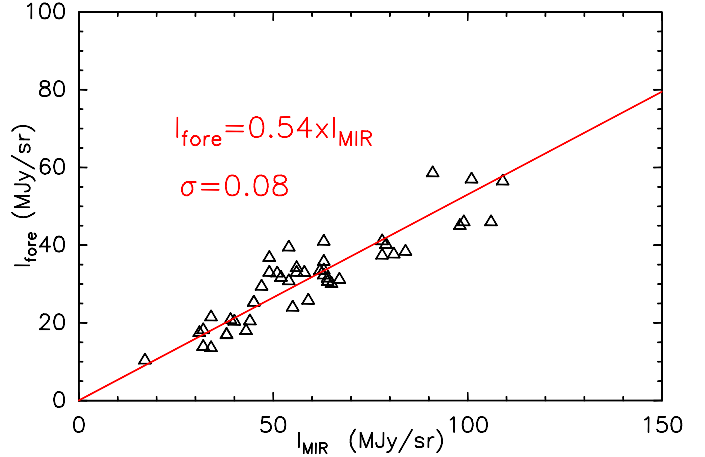}
  \caption{Plot of the 8$\mu$m foreground intensity calculated for
    57 positions (see text) of the \citet{rathborne2006} sample in function of
    the 8$\mu$m mid-infrared radiation field estimated around
    them. The best linear fit is shown as a red solid line.
  \label{ifore-imir}}
\end{figure}

\begin{figure}[!th!]
\includegraphics[width=7.5cm,angle=270]{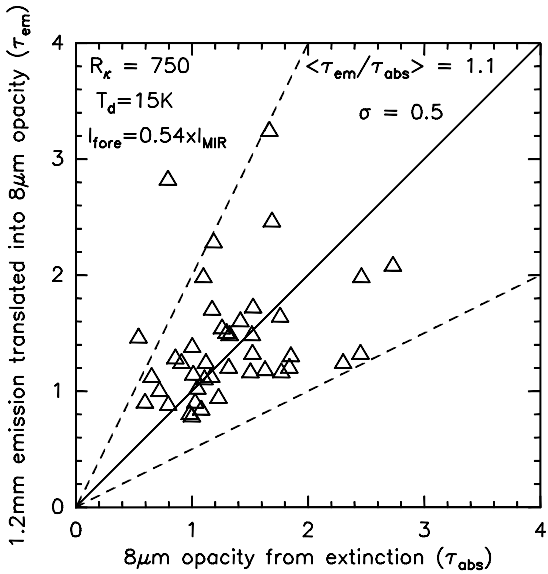}
\caption{Same as Fig.~\ref{rath} but only for starless sources and
  with a 8$\mu$m opacity calculated with $I_{\rm{fore}} =
  0.54\times I_{\rm{min}}$. The solid line marks the relationship $\tau_{\rm{abs}} = \tau_{\rm{em}}$,
  while the two dashed lines indicate $\tau_{\rm{abs}} = 0.5\times\tau_{\rm{em}}$ and
  $\tau_{\rm{abs}} = 2\times\tau_{\rm{em}}$.
   \label{rath_bis}}
\end{figure}

Assuming that for starless cores the true 8$\mu$m opacity is given by
$\tau_{\rm{em}}$, we can invert Eq.~(\ref{eq2}) to estimate the value of
I$_{\rm{fore}}$ in terms of $I_{\rm{MIR}}$. We did such a calculation for
  every starless core and plotted the results in Fig.~\ref{ifore-imir},
  I$_{\rm{MIR}}$ being measured at the position of the core on the large scale
  emission map (Sec.~3.3).  A strong correlation is seen between
I$_{\rm{fore}}$ vs I$_{\rm{MIR}}$.  The best linear fit to this correlation is
given by
 \begin{equation}
 I_{\rm{fore}} = 0.54\times I_{\rm{MIR}}
\label{eqn:fore}
 \end{equation}
 with a standard deviation of 0.08, minimum and maximum values of 0.4 and
 0.75, respectively.
This relationship allows us to compute an average foreground emission
just by estimating the mid-infrared radiation for any IRDCs.
Figure~\ref{rath_bis} shows $\tau_{\rm{em}}$ versus
$\tau_{\rm{abs}}$ calculated using Eq.~(\ref{eqn:fore}), but only for the
starless cores this time. Here $<\tau_{\rm{em}}/\tau_{\rm{abs}}> =1.1$ with a
dispersion of only 0.5.

The relation in Eq.~(\ref{eqn:fore}) gives us the maximum opacity (and
equivalent column density) we can probe before reaching saturation. Indeed,
the rms noise level of the 8$\mu$m images ($\sigma_{\rm{noise}}\sim
0.3$~MJy/sr) defines the minimum flux we can detect above the foreground
emission. Below this value, the dust in the cloud is basically absorbing all
the background emission and we cannot recover the true peak column
density. This saturation opacity, $\tau_{\rm{sat}}$, is given by
$\tau_{\rm{sat}} = -\ln(\sigma_{\rm{noise}}/I_{\rm{bg}})$, with $I_{\rm{bg}} =
0.46\times I_{\rm{MIR}}$. The saturation opacity is calculated for every IRDC
and given in Table \ref{table:1}. We also note that we have $I_{\rm{fore}}
\simeq I_{\rm{bg}}$ as also observed by \citet{johnstone2003} and this
suggests that most of the foreground emission originates from the same place
as the background emission and is local to the IRDC, and therefore the
foreground emission is independent of distance to the IRDC.

\subsection{Construction of the opacity maps}
\label{subsec:opmaps}

To construct opacity maps of IRDCs all over the Galactic plane we
mosaiced the GLIMPSE 8\microns\ and MIPSGAL 24$\mu$m images in blocks
of $1^\circ$ in longitude by $2^\circ$ in latitude using the Montage
software (http://montage.ipac.caltech.edu/). To allow the
identification of IRDCs which cross the edges of these blocks and to
allow the extraction of regions large enough for our analysis around
clouds near the edges of these blocks, each consecutive block overlaps
adjacent blocks by 0.5$^\circ$. In principle this means our extraction
could miss IRDCs larger than about $0.5^\circ$ in size. However the
largest cloud identified by \citet{simon2006a} is 27\arcmin\ long.

The sensitivity of the Spitzer images is such that significant numbers of
stars and galaxies appear in them, even at 8$\mu$m.  These need to be removed
in order to produce clean mid-infrared images and opacity maps of the
clouds. This has been done in two steps. First identifying the central
position of stars in the field using the IDL FIND task from the Astronomy
library. Second, the values in the pixels containing the star were replaced
with values calculated from an average gradient plane fit to the values of the
pixels surrounding the star we want to remove.  While this allowed the
recovery of some part of the structure of a cloud, it can also produce
artifacts.

Once the 8$\mu$m stars were removed, we calculated the mid-infrared radiation
field $I_{\rm{MIR}}$ by smoothing each 8$\mu$m block by a normalised Gaussian
of FWHM=308\arcsec \footnote{this size corresponds to (pixel
  size)$\times2^8$}. This size is a compromise between several parameters: the
typical size of an IRDC, the typical spatial scale of the 8$\mu$m emission of
the Galactic plane and the computation time. Visual inspection of Spitzer
images suggests that most of the clouds are filamentary with a minor axis
which is not larger than a few arcminutes.  The smoothing we have used is well
matched to such clouds and our method will recover their exact structure. For
clouds which are larger than the smoothing length, but which are centrally
condensed, we will detect them but somewhat underestimate their opacity. On
the other hand shallow large clouds will be missed (Section 5 and 6). Using a
larger smoothing length would allow us to better detect these large clouds,
but at the cost of additional processing time and more significantly, the
introduction of spurious artificial clouds, especially where the background
emission is weak.  In any case, distinguishing between a feature due to a
smooth lack of background emission or the presence of a large and low column
density cloud requires observations of tracers in addition to the inferred
mid-infrared extinction. We preferred to convolve the images with a Gaussian
rather than using a median filter in order to better recover potential clouds
adjacent to strong 8$\mu$m emitting structures.

Having calculated $I_{\rm{MIR}}$ we are able to compute both I$_{\rm{fore}}$
and I$_{\rm{bg}}$ images (Section~\ref{sec:ifore}). Then using Eq.~(\ref{eq2})
we can construct the 8$\mu$m opacity image, but before doing so, we smoothed
the 8$\mu$m images with a 4\arcsec\ Gaussian in order to suppress high
frequency noise.

A series of artifacts, and spurious clouds may arise from our method.  The
first one comes from potentially interpreting every decrease in the 8$\mu$m
emission on spatial scale smaller than $\sim 5\arcmin$ as being a potential
cloud.  This effect is especially important at high latitudes where the
mid-infrared radiation field is low. In these regions a small decrease in the
intensity will be interpreted as a stronger increase in the opacity 
than for a similar intensity drop in a high mid-infrared radiation field
environment.  Identifying such spurious clouds is difficult, and only
follow-ups in other tracers in emission will give a definitive answer on the
nature of these sources.  However, we have attempted to minimise such objects
by selecting a relatively high opacity detection threshold.

Another artifact can arise in regions with strong intensity gradients in the
initial 8$\mu$m block where the smoothing may artifially produce features
identified as clouds, although real clouds also exist in these environments
\citep[]{zavagno2009}.
%
To help identify possible spurious objects in regions of large  8\microns
  intensity variations, our catalogue (Table~\ref{table:1})\footnote{The full
  catalogue, including images of all the clouds are available online at:
  \textbf{http://www.irdarkclouds.org} or {\bf
    http://www.manchester.ac.uk/jodrellbank/sdc}} lists $\delta \rm{
    I_{MIR}}$,  the normalised maximum variation of $I_{\rm{MIR}}$ within
  the IRDC and defined as $\delta I_{\rm{MIR}} =
  (I_{\rm{MIR}}^{\rm{max}}-I_{\rm{MIR}}^{\rm{min}})/I_{\rm{MIR}}^{\rm{min}}
  $.
Our experience suggests that clouds with $\delta
I_{\rm{MIR}}>0.5$ have to be treated with caution. These clouds represent
14\% of the total number of IRDCs included in our sample.  Overall, after a
visual inspection of every IRDC and the removal of obviously spurious IRDCs, we
believe that more than 90\% of the catalogued objects are true IRDCs.

The tools to automatically construct the maps were mainly constructed
using IDL packages.


\begin{figure*}[!t!]
\includegraphics[width=6cm,angle=270]{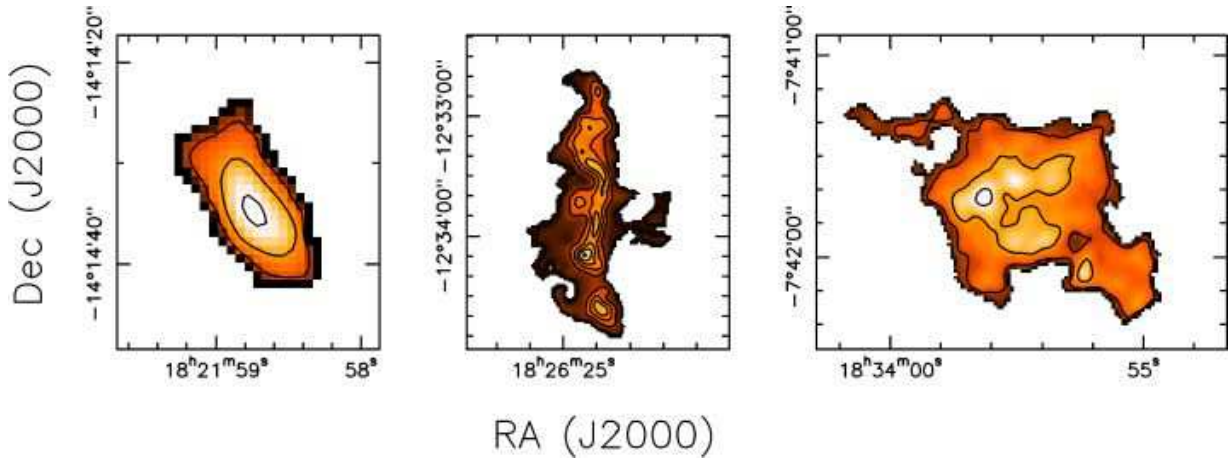}
\caption{ 8$\mu$m opacity maps for the 3 IRDCs 
showed in Fig.~\ref{irdc_samp}. The contours go from 0.4 to 0.8 in steps of
0.2 for the figures on the right and left, while for the middle figure the contours go from 0.4 to to 1.9 in steps of 0.3.
\label{irdc_samp_op}}
\end{figure*}

\section{From 8$\mu$m opacities to column densities}

The images resulting from the analysis described above provide the spatial 8$\mu$m
opacity distribution towards IRDCs. However a more useful quantity is the
H$_2$ column density distribution of these clouds. To convert 8$\mu$m
opacities to H$_2$ column densities requires a knowledge of the properties of
the absorbing dust. Depending on the line of sight and on the structures
observed e.g. diffuse material or dense material, the dust chemical
composition and thus, the dust properties, are different. In dense clouds like
IRDCs, it is believed that dust grains are larger than in the diffuse
interstellar medium due to coagulation and presence of icy mantles on the
grains. This is supported by ISO \citep{lutz1996}, and more recently Spitzer
\citep{indebetouw2005, roman-zuniga2007}, observations which have shown that
towards dense clouds, the extinction cannot be fitted by a single power-law
from the near-IR up to the mid-IR \citep{draine1984}. The recent work has shown
that in dense clouds the extinction decreases from the near infrared to
$\sim5\mu$m and then reaches a plateau up to the silicate absorption band
around 9$\mu$m. This behavior can be reproduced with dust models having
$R_v\simeq5$ \citep{weingartner2001}, implying larger dust grains (compared to
the commonly used value $R_v \simeq 3$ for diffuse interstellar medium).

  For the IRDCs we therefore adopt a value of $A_{8\mu\rm{m}}/A_{v} =
  0.045$ \citep{indebetouw2005,roman-zuniga2007}. To convert to the molecular hydrogen column density,
  $N_{\rm{H_2}}$ we adopt 
\begin{equation}
A_v= 10^{-21} \times N_{\rm{H_2}}
\end{equation}
from \citet{bohlin1978}, although the more recent work by
\citet{draine2003}, based on the observations of \citet{rachford2002},
suggests a 50\% larger column density per magnitude of extinction.  To
account for this, and other uncertainties, the column densities in
this (and subsequent papers), have been calculated from the 8$\mu$m
optical depth adopting the  relation
\begin{equation}
N_{\rm{H_2}} = \tau_{8\mu\rm{m}} \times 3[\pm1] \times10^{22} \rm{cm}^{-2}
\label{coldens}
\end{equation}

\section{Identification of sources
\label{sec:detect}}

Once the opacity maps have been constructed, we need to extract the
information on the structures lying within them. For this purpose, we have
developed a new code, largely inspired by the CLUMPFIND source extraction code
of \citet{williams1994}.  The operation of the code is described in
Appendix~\ref{sec:appt}. The main differences compared to CLUMPFIND are how a
source is defined and its properties determined.  This new method does not
assume that every pixel belongs to a source, but we define the boundaries of
an object by the local minimum between closest neighbours. Then to estimate
the size of the source we calculate the first and second order moments of the
absorption distribution, and then we diagonalise the second order moment matrix
(Appendix~\ref{sec:appt}).

\subsection{IRDCs}
\label{sec:irdcs}

In our maps, the IRDCs have been defined as connected structures lying above
an opacity, $\tau_{8\mu\rm{m}}$, of 0.35 with a peak above 0.7 and a diameter
greater than 4\arcsec.  Therefore, using Eq.~(\ref{coldens}), these detection
thresholds correspond to $1\times10^{22}$~cm$^{-2}$ and
$2\times10^{22}$~cm$^{-2}$, respectively.  With these parameters, we have
identified 11303 IRDCs (see Fig.~\ref{irdc_samp_op}).  Table \ref{table:1}
lists the first 30 IRDCs, giving their name, coordinates, I$_{\rm{min}}$ 
  in MJy/sr, I$_{\rm{MIR}}$ in MJy/sr, $\delta {\rm I_{MIR}}$ (see Sec.~3.3), 
  $\Delta X$ the major axis size in
  arcseconds, $\Delta Y$ the minor axis size in arcseconds, $\alpha$
the position angle  in degrees ( see Appendix~\ref{sec:appt} for an
  exact definition of these parameters), R$_{\rm{eq}}$ the equivalent radius
which corresponds to the radius of a disc having the same area as the IRDC
 in arcseconds, $\tau_{\rm{peak}}$ the 8$\mu$m peak opacity,
$\tau_{\rm{av}}$ the $8\mu$m opacity averaged over the cloud,
$\tau_{\rm{sat}}$ the saturation opacity as described in Section 3.2, the number
of fragments within the IRDC (Sec.~\ref{sec:frags}),
whether there is a 24\microns star in the field/IRDC or not (Sec.~\ref{sec:stars}), and
$\sigma_{\rm{star}}$ the 24$\mu$m stellar density around the IRDC in number of stars per arcminute squared.

\subsection{IRDC fragments}
\label{sec:frags}

Substructures are seen in almost every IRDC map
(Fig.~\ref{irdc_samp_op}). Since column density peaks likely pinpoint the
sites of the formation of the next generation of stars, identifying these
peaks is crucial in identifying the initial conditions of star formation in
IRDCs. We call these substructures identified within the IRDCs
\textit{fragments}.  We prefer this name, rather than for example, cores, as
they have been called in other papers \citep[e.g.][]{rathborne2006}. The term
core has often been used to identify a substructure which forms one star or a
small group of stars and we do not at this stage wish to imply any physical
interpretation of these structures in IRDCs.  Especially since we do not know
the distance of the bulk of the IRDCs,  we cannot infer any physical
  parameters such as the sizes and masses of the fragments/IRDCs.

To extract the IRDC fragments, we apply the same extraction code used
to identify the IRDCs (Appendix~\ref{sec:appt}). We applied different
values of $\tau_{\rm{step}}$ in order to get a comprehensive picture
of the fragmentation in these IRDCs. In total we identified 20000 to
50000 fragments depending on $\tau_{\rm{step}}$ (from 0.1 to
0.35). For each of these fragments we have measured their positions,
sizes, peak and average opacity, and their 24$\mu$m star
association. As an indication of the degree of fragmentation Table
\ref{table:1} includes the number of fragments extracted in each IRDC
with $\tau_{\rm{step}}=0.35$. The nature of these fragments is
discussed in detail in Peretto \& Fuller (2009, in preparation).

\begin{table*}
\caption{Average properties of IRDCs and fragments (extracted with
      $\tau_{\rm{step}}=0.35$). }             
\label{table:2}      
\begin{tabular}{cccccccccccc}        
  \hline
  \hline
  Structures &   Number of  & \multicolumn{2}{c}{R$_{\rm{eq}}$} & \multicolumn{2}{c}{Aspect ratio} &\multicolumn{2}{c}{$\tau_{\rm{av}}$}  & \multicolumn{2}{c}{$\tau_{\rm{peak}}$}  & Star association \\
 & Objects & Average & Range &  Average & Range &  Average & Range &  Average & Range & \\
    & & (arcsec) &(arcsec) & & & &   &  & &$\%$ \\
\hline
   IRDCs  &   11303 &  31 & 4--374 & 2.2 & 1.0--11.6 & 0.15 & 0.01--2.35
   & 1.15 & 0.70 -- 8.36 & 20-68\\
    Fragments  &   19838 &  19 & 1--205  & 2.0 &
    1.0--11.6 & 0.75 & 0.01--7.88 & 1.63 & 0.70 -- 8.36 & 6\\ 
  \hline                                   
\end{tabular}
\end{table*}

\subsection{24$\mu$m point-like sources association}
\label{sec:stars}

In order to check for star formation activity associated with the IRDCs and
fragments, we analysed the 24$\mu$m MIPSGAL data, looking for point-like
sources. For this purpose we used the IDL FIND task of the IDL Astronomy
Library.  As an initial indication of the the star formation activity of these
IRDCs, we have identified all the 24\microns stars lying within a box
(described as {\it Field} in Table \ref{table:1} col. 16) of twice the
calculated extent along the coordinate axes of each IRDC. Doing so, we find
that 32$\%$ of the IRDCs do not have any 24\microns point-like sources in such
a box\footnote{In Table \ref{table:1} columns 16 and 17 \textit{y} stands for
  \textit{yes} and indicates the presence of a star within the field (and/or
  the cloud), while \textit{n} indicates there are no such stars}. On the
other hand, 20$\%$ of the IRDCs have a 24\microns source lying within their
boundaries (Table~\ref{table:1} col. 17). Therefore, the percentage of active
star forming IRDCs is likely to be between 20 and 68$\%$. A more detailed
analysis of the stellar content of IRDCs will be presented in a following
paper.

Concerning the fragments, between 1\% and
6\% have stars lying within their boundaries, depending on the parameters used to
extract the fragments (Peretto \& Fuller 2009, in preparation).

We have also calculated the 24$\mu$m stellar surface density around each IRDC
extracted (Table \ref{table:2} col. 18). This number provides an idea of the
crowding in the area around the IRDC.

\subsection{Uncertainties on the opacity estimates
\label{sec:unc}}

\begin{figure}[!t!]
\includegraphics[width=11.5cm,angle=270]{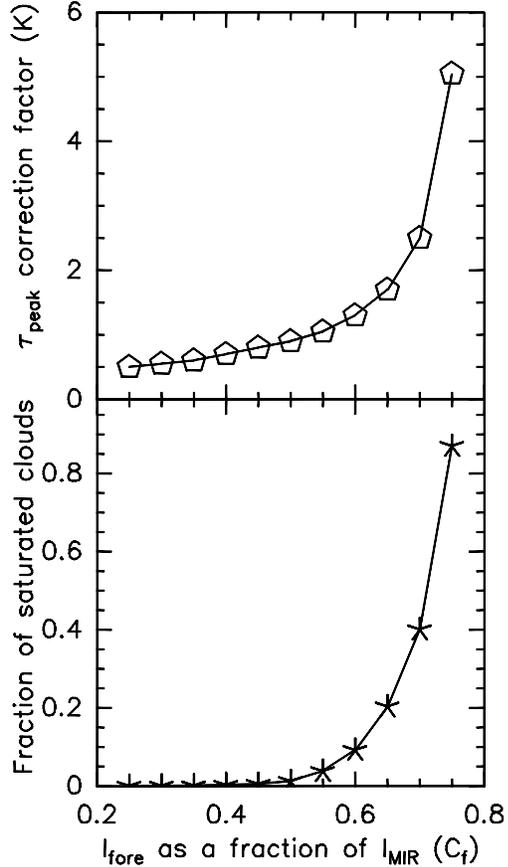} 
\caption{ {\bf (top):} Correction factor to apply to peak opacities in
  order to correct for different foreground intensities than the one
  we used in this study. {\bf (bottom)}: Fraction of saturated clouds
  as a function of the assumption made on the foreground
  intensity.}  \label{bgerr}
\end{figure}

The main source of uncertainty in the opacity maps arises from the
estimate of the foreground intensity $I_{\rm{fore}}$. As explained in
Section 3, we used the relation $I_{\rm{fore}}=0.54\times
I_{\rm{MIR}}$ to calculate this quantity for every cloud. However, as
can be seen in Fig.~\ref{ifore-imir} a dispersion of $\sim0.1$ exists
on this relation with a maximum variation of $\pm0.25$.  To assess the
impact of such variations on the calculated peak opacities of the
clouds we have computed for every cloud the ratio, $K$, of the peak
opacity inferred assuming $I_{\rm{fore}}= C_f\times I_{\rm{MIR}}$
where $0.25<C_f<0.75$ to the peak opacity calculated with the fiducial
$I_{\rm{fore}}$ (Eq.~\ref{eqn:fore}; $C_f=0.54$). Figure~\ref{bgerr}
shows the median value of this ratio as a function of $C_f$.  For each
value of $C_f$ we also calculated the dispersion in $K$ across the
entire sample of clouds. These dispersions were all $< 0.1$, except
for the case $C_f=0.75$ where the dispersion in $K$ reached 0.3.  The
range in $K$ shown on Fig.~\ref{bgerr} provides an estimate of the
peak opacity uncertainty related to the choice/variation of
$I_{\rm{fore}}$. In most cases this uncertainty is less than a factor
of 2, but can be as large as 10 for extreme cases.  On the same
  figure we also plot the fraction of saturated clouds in function of
  the adopted $I_{\rm{fore}}$. Naturally, the higher $I_{\rm{fore}}$,
the higher the number of saturated clouds, reaching 80$\%$ in the most
extreme case, but being less than 10$\%$ for $I_{\rm{fore}} < 0.6
\rm{I}_{\rm{MIR}}$. In the case of $C_f = 0.54$, the percentage of
saturated cloud is $3\%$. This is consistent with a visual inspection of the 8\microns
intensity profiles of a sample of clouds which indicates that less
than 10\% of the objects show a flattening in their inner regions, a
signature of possible saturation.

Another source of uncertainty is the variation of the foreground
intensity relative to the background emission. Since we have shown
that on average the background emission is equal to the foreground
emission (Sec.~\ref{sec:ifore}), we assumed that the variations of
both quantities in front and behind a cloud have the same origin, and
so, the same variations. However, this assumption could be wrong. For
instance one could be constant over the extent of the cloud, more
likely the foreground, with the other one containing all the
variations observed in the mid-infrared radiation field. The impact of
such effects on the opacity estimate is similar to the one described
above. Clouds with small variations in
their mid-infrared radiation fields are thus better constrained than
the ones with high $\delta I_{\rm{MIR}}$.

As mentioned in the previous section large clouds ($> 5\arcmin$) have
opacities which are likely to be underestimated, however this effect is minor
compared with those mentioned above. Overall, considering all the factors
which contribute to the uncertainty in opacity, we estimate the values
derived from the Spitzer data are uncertain by a factor of no more than
two. This result is consistent with the observations of a subset of clouds in
the 1.2mm continuum emission from the dust (Fig.~\ref{rath_bis}).

\section{Comparison with the MSX IRDC catalogue}

\citet{simon2006a} undertook a systematic survey of IRDCs using MSX
data. Their survey covers a larger area of the Galactic plane than
ours due to the smaller coverage of GLIMPSE survey. In total,
\citet{simon2006a} have extracted 6721 clouds between $10\degr <|l| <
65\degr$ and $-1\degr < b < 1\degr$. For the same coverage we
extracted 11303 Spitzer dark clouds, which is roughly twice
as many. However, the detection limits, peak and boundary, in the two
surveys are different, the simple comparison of the numbers of
clouds provides only an incomplete comparison and so a more
  complete comparison has been performed.

\begin{figure*}[!t!]
\includegraphics[width=18.5cm,angle=0]{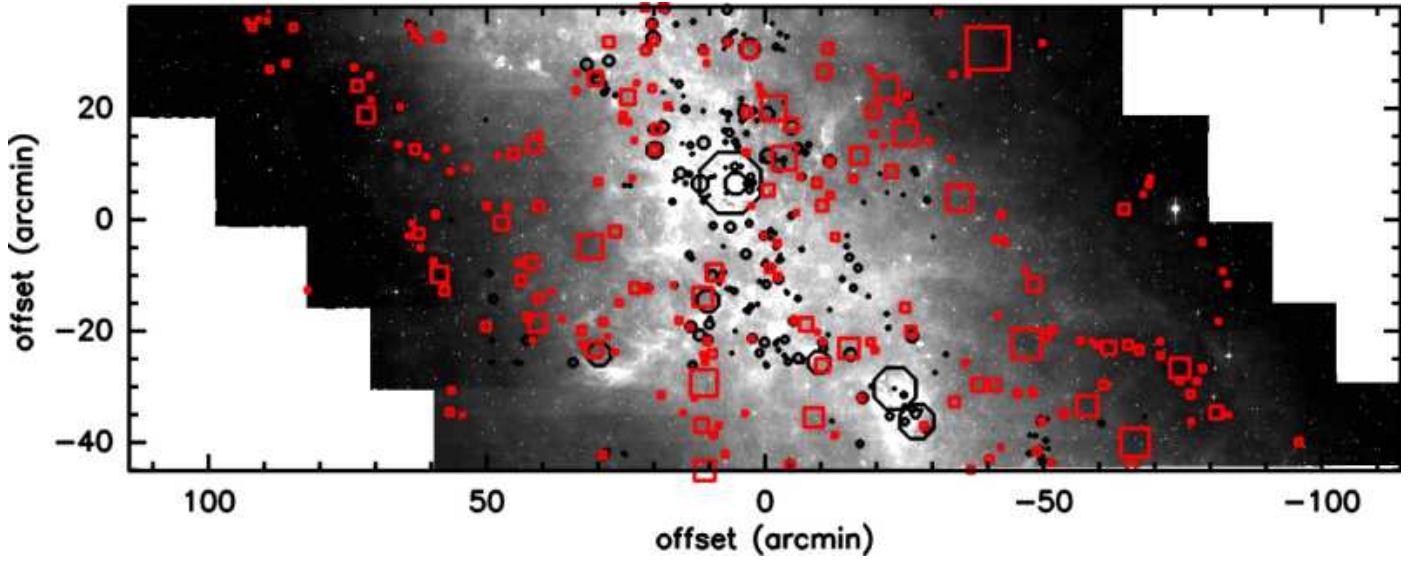} 
\caption{In grey scale is the Spitzer 8$\mu$m emission of one of the
  blocks we constructed around $l\simeq 30\degr$. The black circles
  indicate the position and size of the Spitzer IRDCs identified in
  this study, while the red square symbols code the position and size
  of the MSX IRDCs. We see on this image that the Spitzer IRDCs are
  more numerous where the background is stronger, while, quite
  surprisingly, this is not the case for the MSX IRDCs. The MSX clouds
  detected at $|b|>0.5\degr$, are on average the larger clouds in
  the \citet{simon2006a} sample. For most of them, we do not detect
  any Spitzer IRDCs at these positions in our standard processing
  (using a 5\arcmin\ Gaussian) but some are detected when using a
  larger smoothing function (see text).}
 \label{fig:gl30} 
\end{figure*}

\begin{figure*}[!t!]
\includegraphics[width=6.5cm,angle=270]{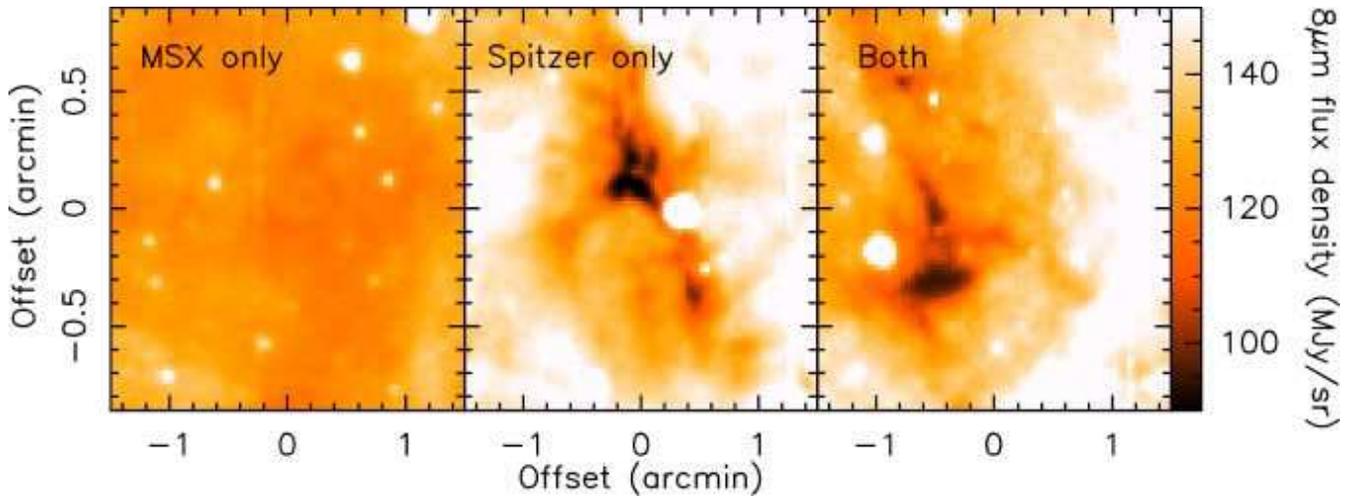} 
\caption{Comparison of three IRDCs seen with Spitzer at 8$\mu$m
  illustrating the 3 categories of IRDC based on their MSX and Spitzer
  detection. Note that the cloud detected only in the MSX catalogue
  (left panel) exhibits much lower extinction than the other two objects.}
 \label{comp_msxsdc} 
\end{figure*}

As illustrated by Fig.~\ref{fig:gl30}, it appears that a minority
  of IRDCs are common to both MSX and Spitzer catalogues. Actually,
  only 20$\%$ of the Spitzer dark clouds appear in the MSX catalogue
  (corresponding to 25$\%$ of MSX clouds being associated with a
  Spitzer dark cloud). Based on this comparison we define 3 categories
  of clouds: {\it Spitzer only}, which are clouds appearing only in our
  catalogue; {\it MSX only}, which are clouds appearing only in Simon et
  al. catalogue; and {\it both}, which are clouds appearing in both
  catalogues. Figure~\ref{comp_msxsdc} shows an example of an IRDC in 
  each of these categories.

\begin{figure}[!t!]
\includegraphics[width=6.5cm,angle=270]{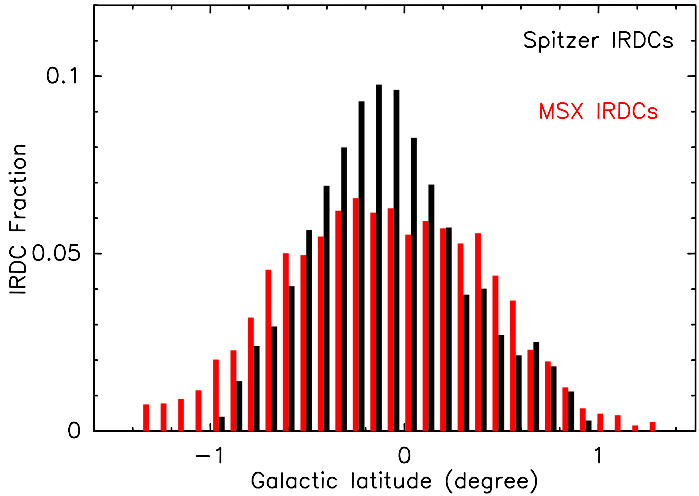} 
\caption{ Comparison of the latitude distribution of Spitzer and MSX
  dark clouds 
  }
 \label{fig:latdist} 
\end{figure}

  Of the {\it Spitzer only} clouds, 51\% do not meet the size criteria, R$_{eq} >
  20\arcsec$, imposed by \citet{simon2006a} to identify the MSX IRDCs,
  explaining why they are not in the MSX catalogue. The remaining
  $\sim30\%$ of {\it Spitzer only} IRDCs are the result from the difference in
  the method used to estimate the background. Using a median filter of
  30\arcmin\ diameter, \citet{simon2006a} underestimated the
  background almost everywhere in the inner $0 < |b| < 0.25\degr$ of
  the Galactic plane. As a consequence, the inferred background
  reaches a similar value as the IRDC itself, and therefore, an IRDC
  is not detected. This artifact can be seen when ploting the source
  fraction as a function of the Galactic latitude
  (Fig.~\ref{fig:latdist}). We see a significant difference between
  the distributions of MSX and Spitzer IRDCs. The MSX IRDCs have a
  rather flat distribution in a central 1\degr\ region whereas the
  Spitzer IRDC distribution has a clear central peak decreasing
  sharply on both sides of it. We believe than this difference
  arises from the difference in the background construction.

  On the other hand the {\it MSX only} clouds  have
  very low contrast (opacity peaks) and are particularly large. The
  detection of such clouds in the MSX data has been possible due to
  the large background smoothing length, and the low contrast
  threshold used by \citet{simon2006a}.  In order to investigate this
  effect and see whether our method could recover these clouds when
  using a larger Gaussian, we smoothed the block shown in
  Fig~\ref{fig:gl30} to 20\arcmin, and performed the extraction of IRDCs
  on the resulting opacity map. Doing so, we find twice as many clouds
  (40\%) which are in both catalogues, but in parallel 35\% of Spitzer
  clouds which were initially detected using a smaller Gaussian are
  lost. The remaining {\it MSX only} clouds are just too shallow to be
  identified given the opacity threshold we used, 0.7. In
  addition, looking at their 8\microns emission it is not clear
  whether many of these clouds are real, or just a decrease in the
  background of the Galactic plane.

  Overall, we can say that 80\% of our catalogue comprises IRDCs which
  were previously unknown and constitutes the most complete catalogue
  available of such objects with column density peaks above
  $1\times10^{22}$~cm$^{-2}$.

\section{Summary}

This paper, the first of a series dedicated to the study of infrared
dark clouds, describes the techniques developed to establish a
complete catalogue of Spitzer dark clouds.  We analysed the full data
set of the 8$\mu$m GLIMPSE Galactic plane to look for IRDCs. We
extracted \nclouds of these clouds, obtaining column density
maps for each of them, and characterizing their physical
properties. We also identify the substructures lying within these
clouds, extracting up to $\sim50000$ of these.
Table~\ref{table:2} presents a summary of the average and range of
properties of both the clouds and these substructures (fragments).
The full table of the properties of the clouds and fragments plus
images and opacity maps are available from an online
database\footnote{The database is available at
  http://www.irdarkclouds.org or
  http://www.manchester.ac.uk/jodrellbank/sdc}.  In subsequent papers
we will exploit the tremendous quantity of information concerning the
initial conditions for the formation of stars in the Galaxy contained
within this set of IRDC column density maps.

\begin{acknowledgements}
  This work was supported in part by the PPARC and STFC grants.  We
  thank Hannah Stacey for her work in the early stages of identifying
  some of the IRDCs. We also thank Jim Jackson, Robert Simon, and Jill
  Rathborne for providing us with the IRAM 30m dust continuum images
  published in \citet{rathborne2006}
This research made use of Montage, funded
  by the National Aeronautics and Space Administration's Earth Science
  Technology Office, Computation Technologies Project, under
  Cooperative Agreement Number NCC5-626 between NASA and the
  California Institute of Technology. Montage is maintained by the
  NASA/IPAC Infrared Science Archive.
\end{acknowledgements}

\appendix

\section{Method for extracting sources}
\label{sec:appt}

\begin{figure}[!t!]
\includegraphics[width=6.cm,angle=270]{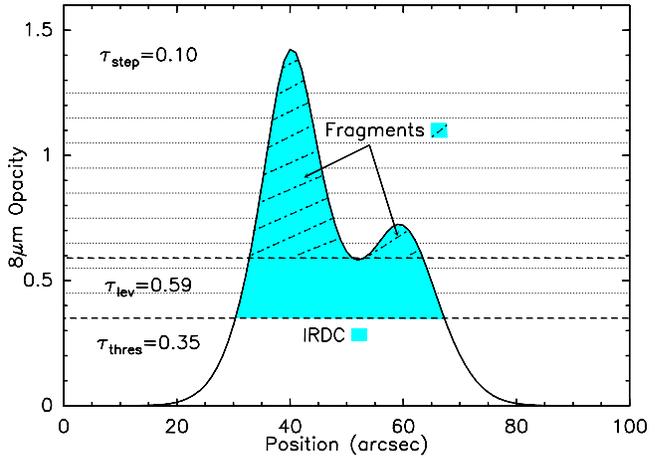}
\caption{Illustration of our extraction method. This figure shows the
  opacity profile of a typical IRDC. The bottom dashed line shows the
  opacity threshold beneath which structures are ignored. 
   The dotted
  lines show the different slices through the cloud, every slice
  being separated by $\tau_{\rm{step}}$. The upper dashed line shows the
  opacity corresponding to the local minimum , $\tau_{\rm{lev}}$, between
  the two local peaks shown on that plot. In such a cloud, our method
  would extract one IRDC (colored area) and two fragments (colored area + dashed-dotted lines) within
  it.  \label{irdc_opprof}}
\end{figure}

\begin{figure}[!t!]
\includegraphics[width=7.5cm,angle=270]{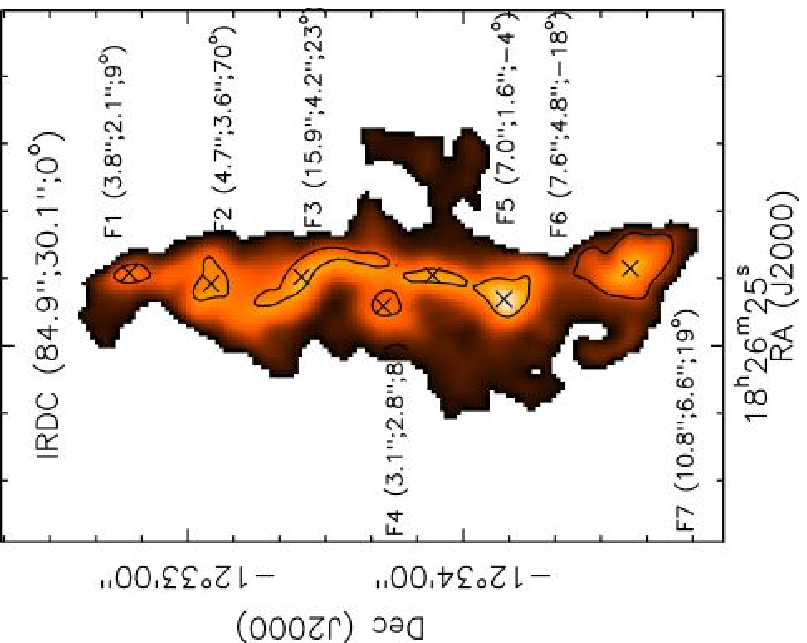}
\caption{8$\mu$m opacity map of the middle IRDC shown in
  Fig.~\ref{irdc_samp}. Our extraction method detected 7 fragments within this
  IRDCs when $\tau_{\rm{step}}=0.1$. The black contours mark the
  $\tau_{\rm{lev}}$ value (boundary contour) for each fragment.  The sizes and
  position angle are also given in between brackets. We can see that these
  values give a reasonable description of the shape of the fragments (and
  IRDC) \label{irdc_op2d}} \end{figure}

We developed a new code to extract sources from our opacity maps. The
first part of our algorithm is mainly based on the same principle as
the one developed by \citet{williams1994} for CLUMPFIND. We set two
main parameters which are the lowest contour level under which we do
not consider any structure, $\tau_{\rm{thres}}$, and a step in unit of
the map,  $\tau_{\rm{step}}$. Then we look at every local peak between
two consecutive levels, up to the maximum of our image. The number of
local peaks gives us the number of fragments we will extract from the
image, unless the final estimated size is lower than the final angular
resolution or that the amplitude between the peak of the fragment and
its external boundary is less than $\tau_{\rm{step}}$. Then we have to
determine the pixels we associate to each local peak. For this, for
every peak, we go down, level by level, and check if the local peak we
are looking at is the only one in this contour. If yes, we look at the
following contour and do the same job. If there is more than one local
peak within the contour we look for the local minimum between these two
peaks, $\tau_{\rm{lev}}$, and the pixels lying above $\tau_{\rm{lev}}$ and
associated with the considered peak define the extent of the fragment.

In order to measure the size of the clouds and fragments, we did not want to
assume any particular shape for the source. So, once we have identify all the
pixels associated with a given peak, we estimate first the center of gravity
of the cores, $(X_{CG},Y_{CG})$, using

\begin{equation}
X_{CG} = \frac{\displaystyle\sum_{i=1}^N V_i \times x_i}{\displaystyle\sum_{i=1}^N V_i} \\
Y_{CG} = \frac{\displaystyle\sum_{i=1}^N V_i \times y_i}{\displaystyle\sum_{i=1}^N V_i}
\end{equation}
where $V_i$ is the value of the $i$th pixel, $x_i$ and $ y_i$ its
coordinates, and N is the number of pixels. Then, we calculate the matrix of moment of inertia, I:
\begin{equation}
I=\left[\begin{array}{cc}
I_{xx}&I_{xy}\\
I_{yx}&I_{yy}\\
\end{array}\right]
\end{equation}

with 
\begin{equation}
I_{xx} = \displaystyle\sum_{i=1}^N V_i (y_i-Y_{CG})^2 \\
\end{equation}

\begin{equation}
I_{yy} = - \displaystyle\sum_{i=1}^N V_i (x_i-X_{CG})^2\\
\end{equation}

\begin{equation}
I_{xy} = I_{yx} = \displaystyle\sum_{i=1}^N V_i (x_i-X_{CG})(y_i-Y_{CG})
\end{equation}

Finally, we diagonalize I in order to obtain its two eigenvalues and
eigenvectors. From this we can easily calculate the position angle
$\alpha$ of the major axis (given by the vector associated by the
smallest eigenvalue). To estimate the sizes of the cores we calculate
the following values:

\begin{equation}
\sigma_X^2 = \displaystyle\sum_{i=1}^N  \left(\left[ x_i \cos(\alpha) - y_i \sin(\alpha) \right] - \left[ X_{CG} \cos(\alpha) - Y_{CG} \sin(\alpha) \right] \right)^2
\end{equation}

\begin{equation} 
\sigma_Y^2 = \displaystyle\sum_{i=1}^N \left(\left[ x_i \sin(\alpha) +
y_i \cos(\alpha) \right] - \left[ X_{CG} \sin(\alpha) + Y_{CG}
\cos(\alpha) \right] \right)^2 
\end{equation}

The sizes are then estimated by $\Delta X =
2\times\sqrt{\sigma_X^2/N}$ and $\Delta Y =
2\times\sqrt{\sigma_Y^2/N}$
  
  The three values, $\Delta X$, $\Delta Y$ and $\alpha$, are given for every
  IRDC  in
  Table~\ref{table:1}.

\bibliographystyle{aa}
\bibliography{references}



\end{document}